\documentclass[12pt]{iopart}
\usepackage{graphicx,amssymb,epsf}

\usepackage{geometry}
\begin{document}
%\review{Generalized Parton Distributions and Deeply Virtual Compton Scattering}
\title{Generalized Parton Distributions and Deeply Virtual Compton Scattering}

\author{Marie Bo\"er, Michel Guidal}
\address{Institut de Physique Nucl\'{e}aire Orsay, CNRS-IN2P3, Universit\'e Paris-Sud, France}

\begin{abstract}
We present a method which allows to extract theoretical 
informations out of a limited set of experimental data and observables, 
forming up in general an under-constrained system. It has
been applied to the field of nucleon structure, in the domain of
Generalized Parton Distributions (GPDs). We take advantage of this 
review to remove a couple of approximations that we used in our 
previous works and update our results using the latest data published.

\end{abstract}
%\pacs{}
\maketitle
%\clearpage
%\tableofcontents 
%\newpage

\section{Introduction}

The Generalized Parton Distributions (GPDs) are functions which 
parametrize several aspects of the complex composite quark and gluon (parton)
structure of the nucleon, which is, up to now, not fully calculable 
from the first principles of Quantum Chromo-Dynamics (QCD). In particular, 
the GPDs allow to determine, in a  frame where the nucleon goes to the 
speed of light in a given direction, the (transverse) spatial distribution
of the partons in the nucleon as a function of their (longitudinal) momentum. 
This allows for a sort of ``tomography" of the nucleon where one probes the transverse
size of the nucleon for different partons' momentum slices. In other words, 
one can follow
how the nucleon size evolves as one probes the valence or sea quark/gluon regions.
In this article, we discuss how to extract several GPD-related quantities from experimental data. 
We introduce in the next section the basic concepts and formalism of GPDs in order
to understand the problematics for readers who are not familiar with the subject and, in the following section, we discuss the actual and numerical extraction of the GPD information from the existing data.
We will conclude in the final section.

\section{Generalized Parton Distributions}

We refer the reader to the reviews~\cite{Goeke:2001tz,Diehl:2003ny,Belitsky:2005qn,
Guidal:2013rya,Boffi:2007yc} for details on the theoretical formalism. We  
summarize in this section a few main points and limit our discussion to quark GPDs and 
to the QCD leading-order formalism, which we will use throughout the remaining of the
article.

\begin{figure}[htb]
\begin{center}
\includegraphics[width =9.cm]{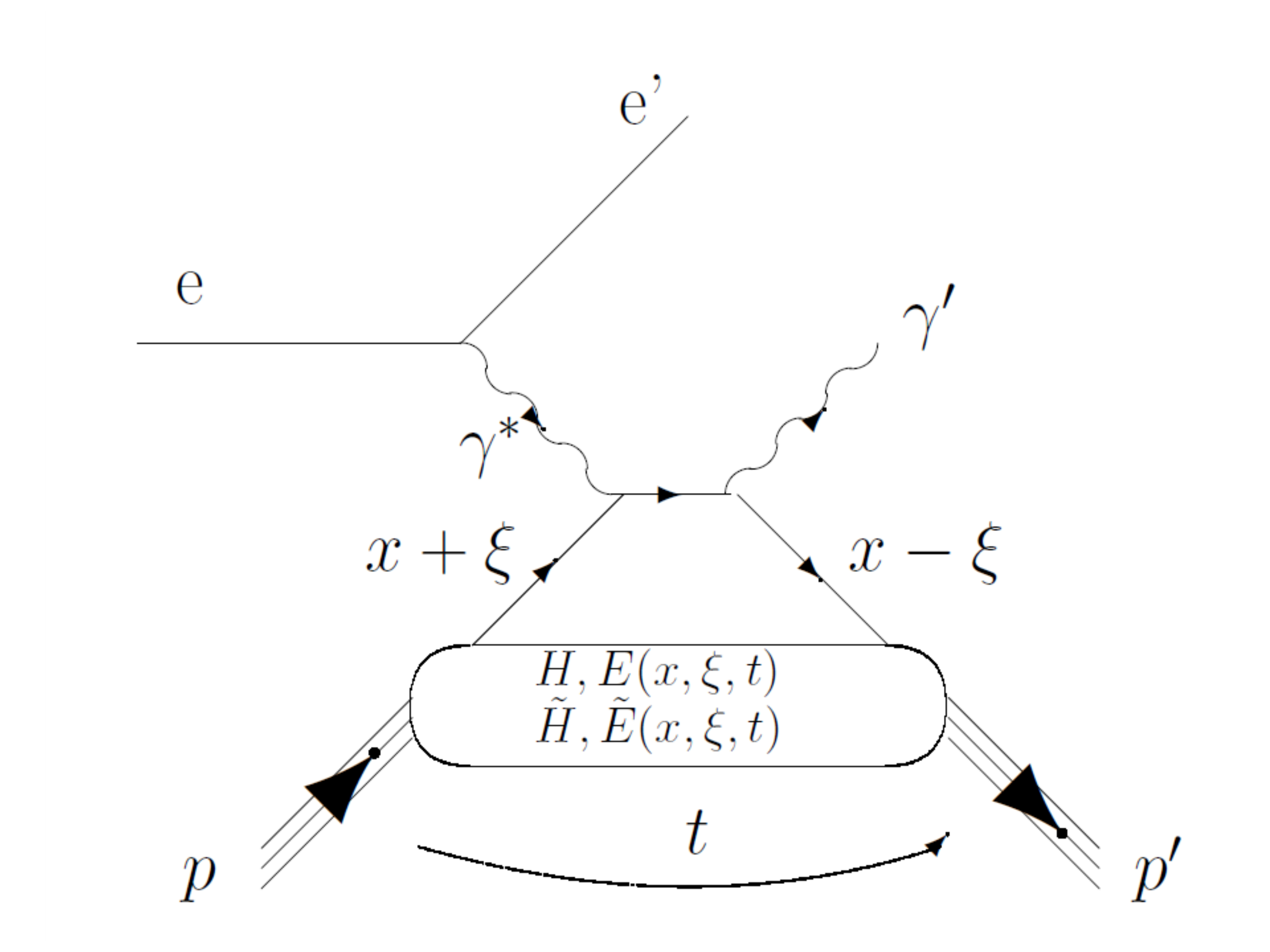}\hspace*{-1cm}
\includegraphics[width =8.cm]{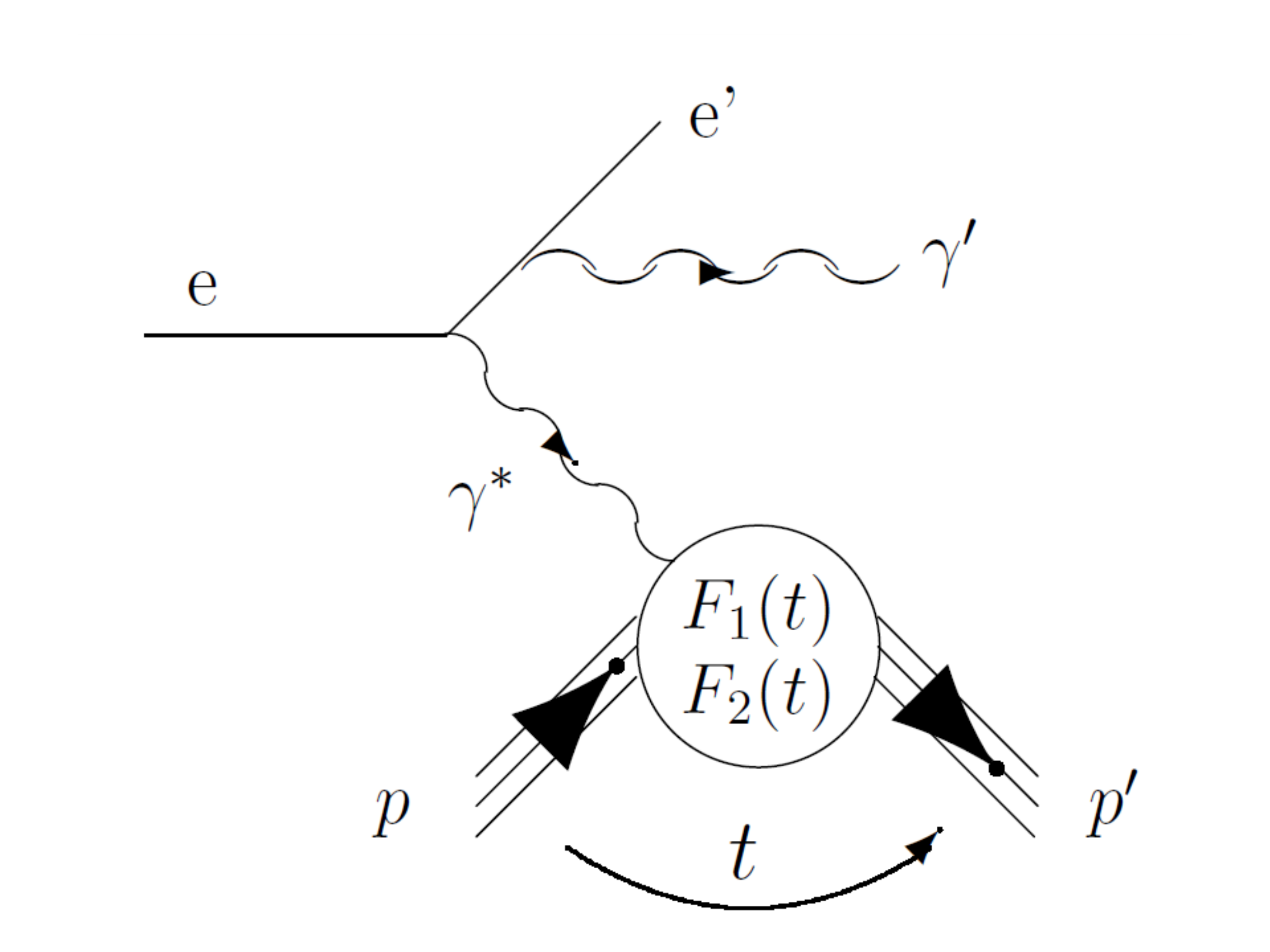}
%\vspace{-6.cm}
\caption{Left: The handbag diagram for the DVCS process on the proton $ep\to e'p'\gamma '$. 
There is also a crossed diagram which is not shown here. The longitudinal momentum
fractions $x+\xi$ and $x-\xi$ of the initial and final quarks are indicated
as well as the total (squared) momentum transfer $t$ between the final and initial nucleon. 
Right: The BH process
also contributing to the $ep\to e'p'\gamma '$ process. The real photon can also be radiated from
the incoming electron.}
\label{fig:dvcs}
\end{center}
\end{figure}

Nucleon GPDs are the most simply accessed in the hard exclusive electroproduction 
(or more generally leptoproduction) of a photon off the nucleon. This process,
$eN\to e'N'\gamma'$, is called Deep Virtual Compton Scattering (DVCS)
and is depicted in Fig.~\ref{fig:dvcs}-left for the proton case. QCD factorization theorems state that, 
at large electron momentum transfer $Q^2=(e'-e)^2$ and small nucleon 
momentum transfer $t=(p-p')^2$, the process in which the \underline{same} quark 
(or antiquark) absorbs the incoming virtual photon and radiates the final real photon
before returning into the nucleon is the dominant one. This is Compton scattering at the quark level and, like at the atomic or molecular level, intuitively, the idea is that the energy and angular distributions of the final photon will
reflect the momentum or/and space distributions of the inner
constituents of the target, i.e. in the present case of the quarks inside the nucleon. 

These 
distributions are parametrized by four GPDs: $H, \tilde H, E$ and $\tilde E$.
They reflect the spin structure of the 
process. There are indeed four independent 
spin-helicity transitions between the initial and final nucleon-quark systems.
At the nucleon level, the nucleon spin can be flipped or not in the process and 
at the quark level, the process can be helicity-dependent or not.
The GPDs  $E, \tilde E$ ($H, \tilde H$) 
account for the transitions which involve (don't involve) nucleon spin flip and $\tilde H,\tilde E$ ($H, E$) 
account for the quark helicity-dependent (helicity-independent) transitions, making up thus four combinations.

At QCD leading-order, the GPDs depend on three independent variables: $x$, $\xi$ and $t$. In simple terms, GPDs represent the probability amplitude of finding a
quark in the nucleon with a longitudinal momentum fraction $x+\xi$
and of putting it back into the nucleon with a different longitudinal momentum
fraction $x-\xi$, plus some transverse momentum ``kick", which is
represented by $t$. For $\xi$=0, the momentum transfer $t$ becomes the conjugate variable of the impact parameter 
$b_\perp$. In this particular limit, GPD$(x,\xi=0,t)$ provides then, via a Fourier-type transform, the correlation between the transverse spatial distribution of quarks ($b_\perp$-dependence) and their longitudinal momentum distribution ($x$-dependence)~\cite{Burkardt:2000za,Diehl:2002he,Ralston:2001xs}. 

GPDs are actually not directly accessible in the DVCS process because only the variables $\xi$ and $t$ are measurable experimentally: $\xi$ is fully defined by detecting the scattered electron ($\xi\approx\frac{x_B}{2-x_B}$ with 
$x_B=\frac{Q^2}{2m_N(E_e-E_{e'})}$) and $t$ by detecting the recoil nucleon or the emitted photon ($t=(p-p')^2=(\gamma^*-\gamma)^2$). The third variable, $x$, is not measurable (intuitively, it is due to the loop 
in the diagram of Fig.~\ref{fig:dvcs}-left which implies an integral over $x$). 

Experimentally, one can therefore access, from DVCS, only the following eight GPD-related quantities which depend only on $\xi$ and $t$ and which are called Compton Form Factors (CFFs)\footnote{Be aware of slightly different vocabulary (``sub-CFFs") and notations 
for these quantities in the litterature: $-\pi$ factors in the definition of the ``$Im$" CFFs,
 ``-" signs for the ``$Re$" CFFs,...}:

\begin{eqnarray}
H_{Re}(\xi , t) &\equiv& {\cal P} \int_0^1 d x \left[ H(x, \xi, t) - H(-x, \xi, t) \right] C^+(x, \xi),\label{eq:eighta} 
\\
E_{Re}(\xi , t) &\equiv&   {\cal P}  \int_0^1 d x \left[ E(x, \xi, t) - E(-x, \xi, t) \right] C^+(x, \xi),\label{eq:eightb} 
\\
\tilde H_{Re}(\xi , t) &\equiv&  {\cal P}  \int_0^1 d x \left[ \tilde H(x, \xi, t) + \tilde H(-x, \xi, t) \right] C^-(x,
\xi),\label{eq:eightc} 
\\
\tilde E_{Re}(\xi , t) &\equiv&  {\cal P}  \int_0^1 d x \left[ \tilde E(x, \xi, t) + \tilde E(-x, \xi, t) \right] C^-(x,
\xi),\label{eq:eightd} 
\\
H_{Im}(\xi , t) &\equiv& H(\xi , \xi, t) - H(- \xi, \xi, t),\label{eq:eighte} \\
E_{Im}(\xi , t) &\equiv& E(\xi , \xi, t) - E(- \xi, \xi, t), \label{eq:eightf} \\
\tilde H_{Im}(\xi , t) &\equiv& \tilde H(\xi , \xi, t) + \tilde H(- \xi, \xi, t), 
%\;\;\;\;\text{and}
\label{eq:eightg} 
\\
\tilde E_{Im} (\xi , t) &\equiv& \tilde E(\xi , \xi, t) + \tilde E(- \xi, \xi, t), \label{eq:eighth} 
\end{eqnarray}
with the coefficient functions $C^\pm$ (which arise from the quark propagator in the diagram of Fig.~\ref{fig:dvcs}-left) being defined as:
\begin{equation}
C^\pm(x, \xi) = \frac{1}{x - \xi} \pm \frac{1}{x + \xi}.
\end{equation}
\noindent

One final complication comes from the fact that DVCS is not the only process
leading to the final state $eN\to e'N'\gamma'$. There is also the Bethe-Heitler (BH) process, 
in which the final state photon is radiated by the electron (incoming or scattered) and 
not by the nucleon itself, and which interferes with DVCS. It is depicted in Fig.~\ref{fig:dvcs}-right. The non-perturbative structure of the nucleon is in this case described
by the nucleon form factors (FF) $F_1(t)$ and $F_2(t)$. These have been extensively
measured experimentally and can be considered as well-known 
at small $t$. The BH process is thus actually precisely calculable.
One should note that the BH process has a singular behavior 
when the radiated photon is emitted in the direction of the (incoming or scattered) 
electron. If we define $\phi$ as the azimuthal angle between the electron scattering 
plane (containing $e$ and $e'$) and the hadronic production plane (containing $\gamma^*$ and $\gamma'$), the BH cross section displays a sharp rise around $\phi$=0$^\circ$, i.e. when the
three final state particles, e', $\gamma'$ and $p'$, are essentially in the same plane.

The amplitudes of the DVCS and BH processes can be calculated from the first principles
of field theory (Quantum ElectroDynamics, for this matter) using the Feynman rules
for the diagrams of Fig.~\ref{fig:dvcs} in terms of, respectively,
CFFs and FFs. In Ref.~\cite{Belitsky:2001ns}, analytical relations linking    
observables and CFFs have been derived for the $e p \to e' p'\gamma'$ (i.e. BH + DVCS) process. Since the four GPDs reflect the different spin/helicity nucleon/quark 
combinations entering the DVCS process, they are going to contribute in different ways to polarization (beam and/or target) observables. Each polarization observable is in general dominated by one (or a few) CFFs. Here are a few examples of such relations in approximate form:

\begin{eqnarray}
&& \Delta\sigma_{LU}\propto \sin\phi \;\;\{F_1 H_{Im}+\xi(F_1+F_2)\tilde{H}_{Im}-kF_2 E_{Im}+...\}\label{eq:relobscffa}\\
&& \Delta\sigma_{UL}\propto \sin\phi \;\;\{F_1\tilde{H}_{Im}
+\xi(F_1+F_2)\left({H}_{Im}+\frac{x_B}{2} E_{Im}\right)
-\xi kF_2\tilde{E}_{Im}+...\}\label{eq:relobscffb}\\
&& \Delta\sigma_{LL}\propto (A+B\cos\phi) \;\;\left\{F_1\tilde{H}_{Re}
+\xi(F_1+F_2)\left({H}_{Re}+\frac{x_B}{2}E_{Re}\right)+...\right\}\label{eq:relobscffc}\\
&& \Delta\sigma_{Ux}\propto \sin\phi \;\;\left\{k(F_2 H_{Im}-F_1 E_{Im})+...\right\}\label{eq:relobscffd}
\end{eqnarray}
where $\Delta\sigma$ stands for a difference of polarized cross sections,
with the first index referring to the polarization
of the beam: ``U" for unpolarized and ``L" for longitudinally polarized; and the second one 
to the polarization of the target: ``U" for unpolarized, ``L" for longitudinally polarized and ``x" or ``y" for transversely polarized  (``x" is in the hadronic plane and
``y" is perpendicular to it). The kinematical variable $k$ is defined as~: $k= -t / (4m_N^2)$.  In these approximate formula, there is always the product of a FF and a CFF, which reflects the interference of the BH and DVCS processes. Each observable has a dominant harmonic ($\phi$-)dependence, i.e. a $\sin\phi$, $\cos\phi$ and/or a constant. In a first approximation, neglecting terms multiplied by kinematical factors such as
$\xi$, $x_B$ and $k$, one can see that, on a proton target, $\Delta\sigma_{LU}$ ($\Delta\sigma_{UL}$, $\Delta\sigma_{LL}$, $\Delta\sigma_{Ux}$,...)  is dominantly
sensitive to $H_{Im}$ ($\tilde H_{Im}$, $\tilde H_{Re}$, $H_{Im}$ and $E_{Im}$, ...) and shall exhibit a dominant sin$\phi$ ($\sin\phi$, constant + $\cos\phi$, $\sin\phi$, ...)-dependence.

In summary, it is clearly a non-trivial task to extract the GPDs from the experimental data:
there are four GPDs, they depend on three variables, out of which only two are
experimentally measurable, and there is the BH process interfering with the DVCS process 
(which, although rather precisely calculable, exhibits strong variations in some parts of the phase space). Extracting the GPDs is therefore a challenging long-term task. 
It involves, on the one hand, an important experimental program aiming at measuring a series of polarized (and unpolarized) observables for the $ep\to ep\gamma$ reaction over a ($\xi,t$) as broad as possible and, on the other hand, a global theoretical and phenomenological
analysis effort to extract, in a first step, the CFFs from these observables.
In particular, the interfering BH contribution must be deconvoluted. Ultimately, 
the $x$-dependence of the GPDs which is not directly accessible has also to be unraveled
(this can be done only in a model-dependent way for DVCS). 

Several strategies are currently being developped for this long path. 
In the following section, we will describe a method which has been developped in Refs.~\cite{Guidal:2008ie,Guidal:2009aa,Guidal:2010ig,Guidal:2010de} 
and which consists, as an intermediate and first step, in extracting the CFFs from a given set 
of DVCS observables at a given $(\xi, t)$ point. The ultimate goal is of course to 
access GPDs, not simply CFFs, for which the $x$-dependence has still to be resolved.
However, extracting the CFFs, that depend only on two
variables which can be both measured experimentally, has the merit to 
already deconvolute the BH contribution, which can be 
done in an almost model-independent way (provided one has enough constraints, 
i.e. experimental observables, to fit, as we shall see). It also appears 
that some actual physics, i.e. information on nucleon structure, can be infered,
modulo a few reasonable assumptions and corrections, from the CFFs, which we will briefly mention in conclusion.

\section{From data to Compton Form Factors}

The currently available $ep\to e'p'\gamma'$ data which potentially lend themselves to a GPD interpretation, i.e. in a first approach with the requirement $Q^2>$ 1 GeV$^2$, are rather scarce. They have been collected at three experimental facilities: JLab Hall A, CLAS (JLab Hall B) and HERMES. They consist of:

\begin{itemize}
\item the $\phi$-dependence of the beam-polarized
  and unpolarized cross sections~\cite{Munoz Camacho:2006hx},
  at $<x_B>\approx0.36$, and for four $t$ values, 
  at a 
  beam energy of $\approx$ 5.75 GeV, measured by the JLab Hall A collaboration, 
\item the $\phi$-dependence and $\phi$-moments of, respectively, the beam-polarized and longitudinally polarized target spin
  asymmetries\footnote{Here, and in the following, we call ``asymmetries" the ratio of the difference of polarized cross sections to the unpolarized one, i.e. the asymetries are of the form $\Delta\sigma / \sigma$.
We follow the usage in the literature although it would be more appropriate to speak of ``relative asymmetries".}~\cite{Girod:2007aa,Chen:2006na} in the range
  $0.11<x_B<0.58$, and for several $t$ values, at a beam energy of $\approx$ 5.75 GeV,
  measured by the CLAS collaboration, 
\item the $\phi$-moments of the beam spin asymmetries~\cite{Airapetian:2001yk,Airapetian:2012mq,Airapetian:2012pg},
longitudinally polarized target asymmetries~\cite{Airapetian:2010ab}, 
transversally polarized target asymmetries~\cite{Airapetian:2008aa,Airapetian:2011uq}, 
beam charge asymmetries~\cite{Airapetian:2006zr,Airapetian:2009aa,Airapetian:2009bi} 
and the associated beam spin/target spin and spin/beam-charge double 
asymmetries at $<x_B>\approx 0.09$, and for several $t$ values, 
 at a beam energy of $\approx$ 27 GeV, measured by the HERMES collaboration.

\end{itemize}

In the approach of Refs.~\cite{Guidal:2008ie,Guidal:2009aa,Guidal:2010ig,Guidal:2010de}, the idea is to fit all the observables available at a given ($x_B,t$) point (or
equivalently ($\xi,t$) point), taking as free parameters the eight CFFs 
(Eqs.~\ref{eq:eighta}-\ref{eq:eighth}). The full kinematics (beam energy, $Q^2$, $x_B$, $t$, $\phi$) of the reaction being experimentally well defined and the amplitudes of the DVCS and BH processes being well known theoretically, the only unknowns are these eight CFFs which depend only on $x_B$ and $t$ (the FFs which enter the BH amplitude being considered as known 
as mentioned earlier).

We are going to illustrate the method on the JLab Hall A data, which actually 
make up the most unfavorable case of the three data sets: we have only two constraints, 
i.e. two observables (the beam-polarized $\Delta\sigma_{LU}$ 
and unpolarized $\sigma$ cross sections) and we have to determine eight free parameters, 
i.e. the eight CFFs. 
Furthermore, in the unpolarized cross section, the CFFs, which enter 
at the amplitude level in the DVCS process, come in bilinear combinations.
This is different for the beam-polarized cross section where, due 
to the interference with the BH process, they enter in a linear way 
(see Eq.~\ref{eq:relobscffa}). Our system 
is clearly underconstrained and, in addition, not linear. In comparison, for the CLAS data, we also have only two observables, namely the beam-polarized and longitudinally polarized target spin asymmetries, but the denominator of these asymmetries is dominated by the BH process and therefore, in a first approximation, essentially only the numerator, i.e. the difference of polarized cross sections, is sensitive to the CFFs. The problem still remains under-constrained of course but it is at least almost linear in the CFFs.

In fitting two observables with eight free parameters, there is in principle an infinity of solutions. In these conditions, the JLab Hall A beam-polarized and unpolarized cross sections can certainly be fitted with high quality but many combinations of the eight CFFs can provide an equally good fit and not much shall really be learned this way. The stratagem 
in order to extract real information is then to limit 
the domain of variation of the 8 CFFs to some (8-dimensional) hypervolume,
which has physically well-motivated and/or conservative boundaries. Then, if some observables are dominated by some specific CFFs, a convergence can begin to appear for these specific ``dominant" CFFs.
 
To take a simple and illustrative example, if one has
to solve the following system for $x$ and $y$:
\begin{eqnarray}
3 = y + 0.001 x,
\end{eqnarray}
where there are 2 unknowns and 1 constraint, one has clearly an infinite number of 
solutions if there is no other input
in the problem. However, if one now imposes an external 
constraint, like $x$ has to be confined within some range, say $[-10,10]$, then one can extract the approximate solution:
\begin{eqnarray}
3 = y \pm 0.01 \textrm{    (or     }  y = 3 \pm 0.01) 
\end{eqnarray}
The error on the ``dominant" $y$ variable reflects then the influence of the ``sub-dominant" $x$ variable or the correlation between the two variables. This influence, and therefore the error on $y$, depends on the factor which suppresses the $x$ variable (i.e. $0.001$) and on the allowed variation range ($[-10,10]$).

\begin{figure}
\begin{center}\vspace{-15.cm}
\includegraphics[width =20.cm,height=38cm]{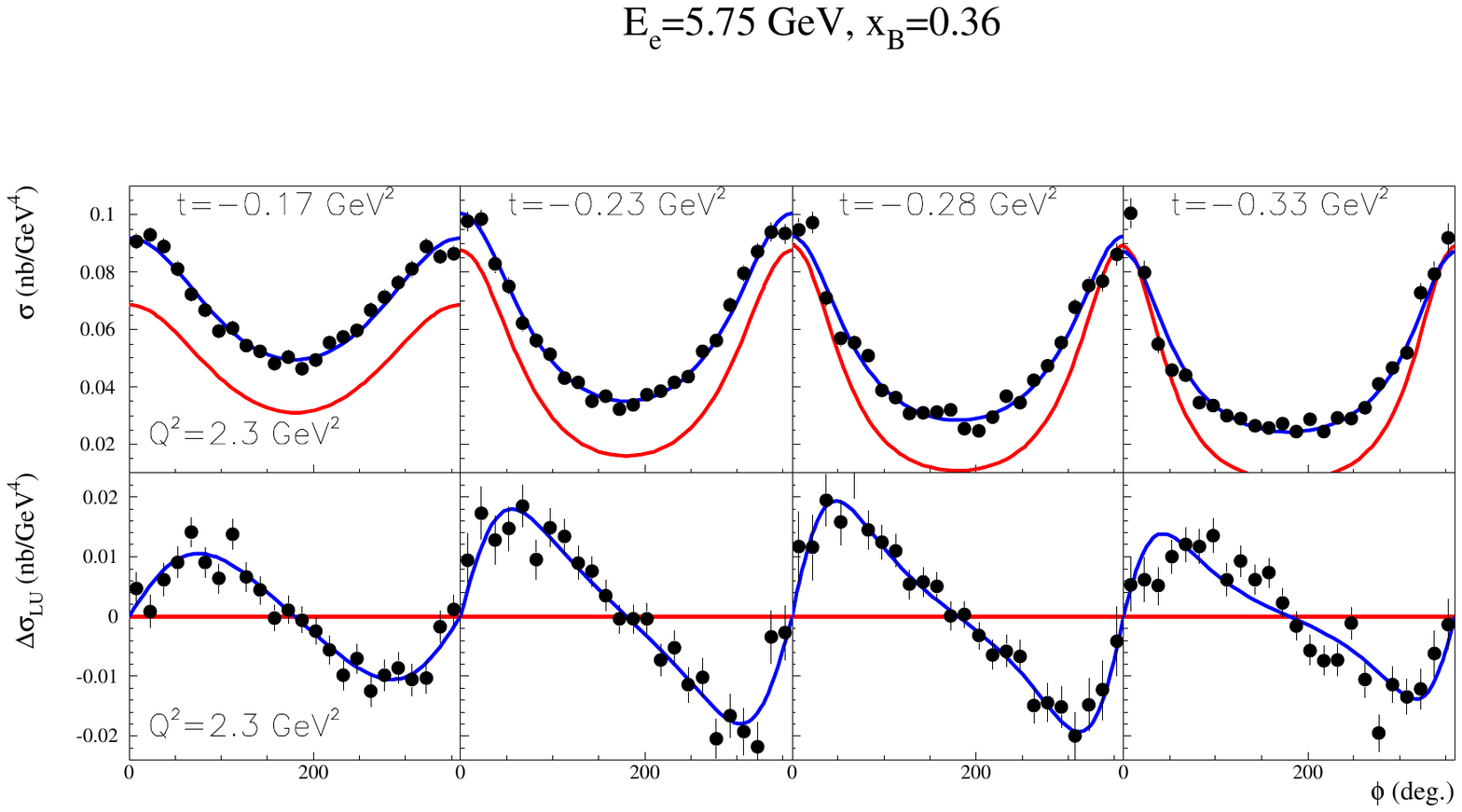}
\end{center}
\vspace{-10.cm}
\caption{Top row: the $ep\to e'p'\gamma'$ unpolarized cross section $\sigma$, and bottom row: the $ep\to e'p'\gamma'$ beam-polarized cross section $\Delta\sigma_{LU}$ 
as a function of $\phi$, at $x_B$=0.36 and $Q^2$=2.3 GeV$^2$ 
for four different $t$ values. The data are from the JLab Hall A collaboration~\cite{Munoz Camacho:2006hx}. 
The blue curves are the results of the fit from our code to the
$\sigma$ and $\Delta\sigma_{LU}$ observables. The red curves show the calculation of the BH alone.}
\label{fig:halla_fit}
\end{figure}

In the case of the Hall A data, according to Eq.~\ref{eq:relobscffa}, the beam-polarized cross section $\Delta\sigma_{LU}$  
should be dominated by the $H_{Im}$ CFF and the unpolarized cross section by the $H_{Re}$ 
CFF (see Ref.~\cite{Belitsky:2001ns}). The hope is thus to determine from the Hall A data some limits
on the $H_{Im}$ and $H_{Re}$ CFFs.

In practice, what we will do is to fit the $\phi$-dependence of the JLab Hall A beam-polarized and unpolarized cross sections. There is a strong $\phi$-dependence of these two observables (see Fig.~\ref{fig:halla_fit}). The peculiar shape of the unpolarized cross section that peaks at $\phi$=0$^\circ$ (or 360$^\circ$) arises from the BH singularities that we already mentioned. The (model-independent) calculation of the BH alone is actually shown by the red curves in Fig.~\ref{fig:halla_fit}. The CFF (or GPD) effect therefore lies in the difference between the red curve and the data. The beam-polarized cross section $\Delta\sigma_{LU}$ exhibits the expected $\sin\phi$-like shape predicted by Eq.~\ref{eq:relobscffa} and the CFF effect lies in its amplitude (the BH alone doesn't produce any $\Delta\sigma_{LU}$ signal). We recall that the CFFs themselves don't depend on $\phi$ (nor on the beam energy and on $Q^2$ in our framework) and, therefore, for given ($x_B$,$t$) values, we have 24 data points for the unpolarized cross section and 24 for the beam-polarized cross section to fit simultaneously. We use a standard least square method where we minimize $\chi^2$: 

\begin{equation}
\chi^2=\sum_{i=1}^{n}
\frac{(\sigma^{theo}_i-\sigma^{exp}_i)^2}{(\delta\sigma^{exp}_i)^2}
\label{eq:chi2}
\end{equation}

\noindent where $\sigma^{theo}$ is the theoretical DVCS+BH cross section (or difference of cross sections), which depends on the CFFs (the free parameters), $\sigma^{exp}$ is the experimental value of the data and $\delta\sigma^{exp}$ 
is the experimental error bar. The index $i$ runs over all the 2 x 24 $\phi$-points available at a given ($x_B$,$t$) value. 
We use the well-known MINUIT code with the MINOS option. The latter option allows to explore 
in a systematic way the full 8-dimensional 
phase space of the free parameters, step by step, reducing therefore the risk of falling into a local minimum. In addition, it allows to determine reliably the errors on the fitting 
parameters corresponding to $\Delta\chi^2=+1$ above the best-fit point. This is crucial when the problem is not linear and when the $\chi^2$ shape is not a simple parabola, as it is in our case.

Regarding the limiting domain for the variation of the CFFs, we have taken an hypervolume defined by, as a first step, $\pm$5 times the values of some ``reference" CFFs. The ``reference" CFFs that we take are those of the VGG model~\cite{Vanderhaeghen:1998uc,Vanderhaeghen:1999xj,Goeke:2001tz,Guidal:2004nd}. It is a well-known and well-used model which has been available for $\approx$ 15 years. GPDs have to satisfy a certain number of normalization constraints and these are all fulfilled by the VGG model. In general, the model gives the general trends of the (few) existing data. Thus, $\pm$5 times the VGG CFFs make up very conservative bounds. In this approach, it is of course more appropriate to fit, actually not the CFFs themselves, but the ``multipliers"
of the VGG reference CFFs, i.e. eight numbers ranging from $-5$ to $+5$. These ``multipliers" are the ratios
of the fitted CFFs to the reference VGG CFFs. We actually mention at this stage that, for the present work, we have improved on the previous work of Refs.~\cite{Guidal:2008ie,Guidal:2009aa,Guidal:2010ig,Guidal:2010de}, where we had neglected the $E_{Im}$ CFF and where we were actually doing fits with only 7 CFFs. We have removed this approximation in the present work and we are doing a 8-free parameter fit (the range of variation of the $E_{Im}$ CFF is taken the same as for the $E_{Re}$ CFF). Another novel feature that we have implemented in the present study is the inclusion of a few ``higher-twist" effects in the DVCS amplitude: QED gauge-restoration terms and exact kinematics, resulting in correction terms of the order of ${t}\over{Q^2}$. % (see Ref. \cite{Vanderhaeghen:1999xj}).
 The final results are actually only slightly affected by these improvements/changes.

\begin{table*}[htb]
\begin{center}
\begin{tabular}{||l||c|c|c|c||}
\hline
%\hspace*{1.cm}
&-t=0.17 GeV$^2$ &-t=0.23 GeV$^2$& -t=0.28 GeV$^2$& -t=0.33 GeV$^2$ \\
\hline \hline
$a(H_{Im})$ & 1.300 & 1.297 & 1.311 & 1.317 \\
$\sigma^+_{a(H_{Im})}$ & 0.140 & 0.152 & 0.118 & 0.037 \\
$\sigma^-_{a(H_{Im})}$ & -1.122 & -1.103 & -1.031 & -0.269 \\
$a(H_{Re})$ & & & 1.867 & 3.996\\
$\sigma^+_{a(H_{Re})}$ & & & 3.031 & 0.644 \\
$\sigma^-_{a(H_{Re})}$ & & & -0.872 & -0.662 \\
$(H_{Im})_{VGG}$ & 1.907 & 1.712 & 1.567 & 1.436 \\
$(H_{Re})_{VGG}$ & & & 0.588 & 0. 758 \\
$\chi^2$ & 46.3594 & 42.3177 & 66.2582 & 106.4390 \\
\hline
\end{tabular}
\caption{Fitted GPD multipliers $a(H_{Im})$ and $a(H_{Re})$ and their negative 
($\sigma^-_{a()}$) and positive ($\sigma^+_{a()}$) uncertainties resulting from 
the fit of the JLab Hall A $\sigma$ and
$\Delta\sigma_{LU}$ observables. The reference VGG CFFs and the (unnormalized) $\chi^2$ value 
for the best fits are presented in the last three rows.}
\label{tab:fitresult}
\end{center}
\end{table*}

Fig.~\ref{fig:halla_fit} shows the results of our best fit (blue curves) to the unpolarized and beam-polarized cross sections of Hall A. It is not particularly challenging to fit two observables with 8 free parameters. However, what is remarkable is that, among the 8 free parameters, there is a couple of them which come out with well-localized values and well-defined error bars (corresponding to $\Delta\chi^2=+1$). These are, as expected, the $H_{Im}$ and $H_{Re}$ CFFs. 
We list in Table~\ref{tab:fitresult} 
our results for the two multipliers $a(H_{Im})$ and $a(H_{Re})$ for the four Hall A $t$-bins, along with their (asymmetric) errors $\sigma^+$ and
$\sigma^-$ as determined 
by MINOS and the (unnormalized) $\chi^2$ values of the fits.
The $H_{Re}$ CFF actually comes out only for the last two $t$-bins. For the other two $t$-bins, the fit was not able to determine finite error bars. The values for $a(H_{Im})$ which are all around 1.3 mean that the VGG model might underestimate this CFF by about 30\% compared to what the data tell. This has to be nuanced by the fact that the errors are very asymmetric, in particular with a large negative error bar, which might lower the ``true" central value. We also indicate in the table the values of the VGG CFFs which allow, by multiplication with the fitted multipliers, to obtain the measured $H_{Im}$ and $H_{Re}$ CFFs.

\begin{figure}\vspace{-5.cm}
\includegraphics[width =18.cm,height=28cm]{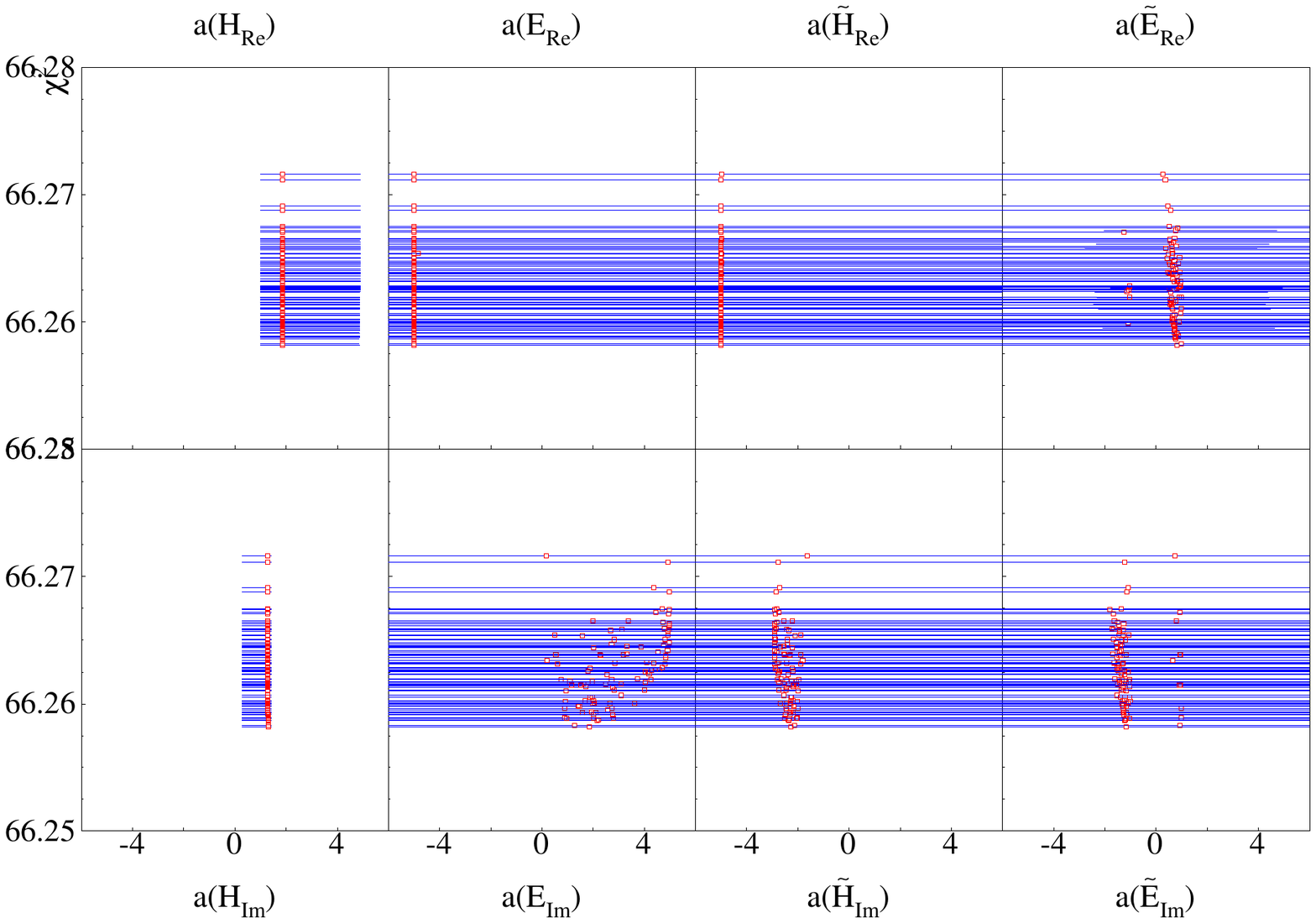}
\vspace{-10.cm}
\caption{Result of the fits for the 8 CFF multipliers $a(...)$ (central value in red and error bar in blue on the $x$-axis) 
and the corresponding $\chi^2$ value ($y$-axis), for the first tens of several hundreds of trials differing only by 
the starting values of the 8 CFF multipliers. This example is for the third $t$-bin of the JLab Hall A data.}
\label{fig:chi2_aall}
\end{figure}

To make sure of the stability and robustness of our fitting procedure, we have run our fitting code several hundreds of times
with a random generation, within the hypervolume defined by $\pm$ 5 times the VGG CFFs, 
of the starting values of the 8 free parameters. Indeed, although MINOS allows to probe the full 8-dimensional hypervolume, 
it uses finite/discrete steps and thus the initial (8-dimensional) starting point might have an influence on the 
final result. Fig.~\ref{fig:chi2_aall} presents, for the third $t$-bin ($-t$=0.28 GeV$^2$), the results of this sampling for the 8 CFFs. The figure shows, for the first few tens of trials, the central values of the 8 multipliers (red dots), their associated error bars (blue bars) and their $\chi^2$, resulting from the fits. We see that the results are different for each of the trials,
either in the $\chi^2$ value or in the central value or in the error bar. This is due to the different starting values of the 8 CFFs. However, we observe that:

\begin{itemize}
\item the difference in the $\chi^2$ values between all the trials is actually
 of the order of the permil, all $\chi^2$ value being, for this $t$ bin, basically comprised between $\approx$ 66.255 and $\approx$ 66.275. Since we have 8 free parameters and we fitted 48 data points, this corresponds to a normalized $\chi^2$ of $\approx$ 1.37.
\item there are clearly two CFFs, namely $H_{Im}$ and $H_{Re}$, which come out systematically with well-defined and almost constant central value and error bars, irrespective of the starting values of the fit. The 6 other CFFs either have widely varying central values or central values at the edge of the domain defined by $\pm$5 times the VGG CFFs and/or ``infinite" error bars (i.e. $\Delta\chi^2=+1$ leads to values outside the domain defined by $\pm$ 5 times the VGG CFFs). For these CFFs, this can be interpreted as the whole range of values within the domain defined by $\pm$ 5 times the VGG CFFs can produce a fit with almost equally 
good $\chi^2$. In other words, these CFFs are essentially 
unconstrained and therefore no particular confidence and meaning can be given to their numerical values. 
\item the error bars on $H_{Im}$ and $H_{Re}$ are very asymmetric: $\approx$ 10\% for the positive error bar and $\approx$ 80\% 
for the negative one for $H_{Im}$ and vice-versa for $H_{Re}$. This reflects the non-linearity and under-constrained nature 
of the problem. The error bars that are obtained reflect actually not the statistical accuracy of the data (which are 
precise at the few percent level as can be seen from Fig.~\ref{fig:halla_fit}) but they are correlation error bars between 
the fitted parameters, as we discussed earlier. The error bars reflect the influence of the CFFs which are not converging, 
in the limit of their variation in the bounded hypervolume defined by $\pm$ 5 times the VGG CFFs.
\end{itemize}

\begin{figure}\vspace{-6.cm}
\includegraphics[width =17.cm,height=33cm]{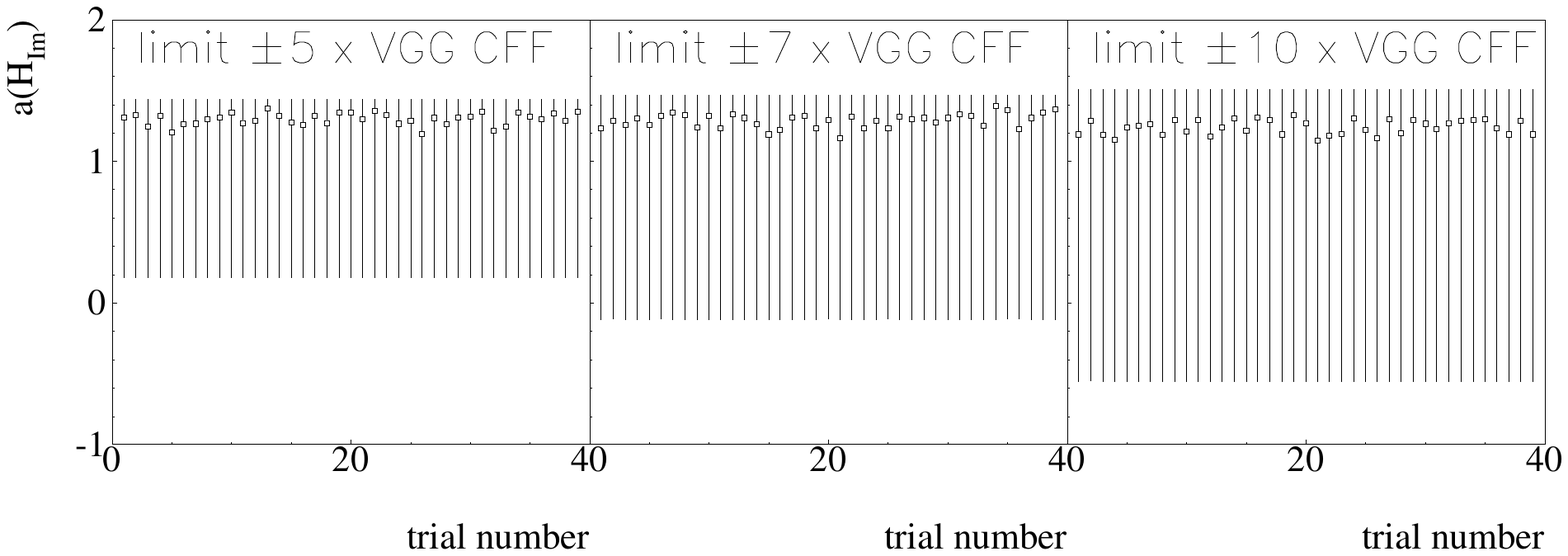}
\vspace{-18.cm}
\caption{Result of the fit for the CFF multiplier $a(H_{Im})$ and its error bar ($y$-axis) for the first 40 of several hundreds of trials ($x$-axis) differing only by the starting values of the 8 CFF multipliers. Left: the domain of variation allowed for the CFFs is $\pm$5 times the VGG CFFs; middle: $\pm$7 times the VGG CFFs; right: $\pm$10 times the VGG CFFs.}
\label{fig:maxrange}
\end{figure}

We actually show in Fig.~\ref{fig:maxrange} the influence of this boundary domain for this same third $t$-bin. Of course, the larger the domain of variation allowed for the CFFs (i.e. $\pm$7 or $\pm$10 times compared to $\pm$5 times the VGG CFFs reference values), the larger the influence of the ``sub-dominant" CFFs on the ``dominant" ones and the larger the error bars of the latter ones. The large error bars presently obtained might not be very spectacular but we recall that we are facing a largely under-constrained problem
with only two observables at our disposal, among which, one contains CFFs in bilinear combinations. We stress that the present approach is essentially model-independent. The only ``external" input is the bounding of the domain of variation of the CFFs by, we deem, very conservative
limits, given that GPDs and CFFs have to obey some (model-independent) normalization constraints. We consider the present approach as minimally model-biased. It is clear that adding new observables to the 8-CFF fit, in particular sensitive to other CFFs than $H_{Im}$, will strongly improve these results. Alternatively, if, guided by some theoretical considerations, one can reduce the range of variation of some of the CFFs into an hyperspace smaller than $\pm$ 5 times the VGG CFFs, the error bars can also obviously only diminish.

\begin{figure}\vspace{-2.cm}
\includegraphics[width =17.cm]{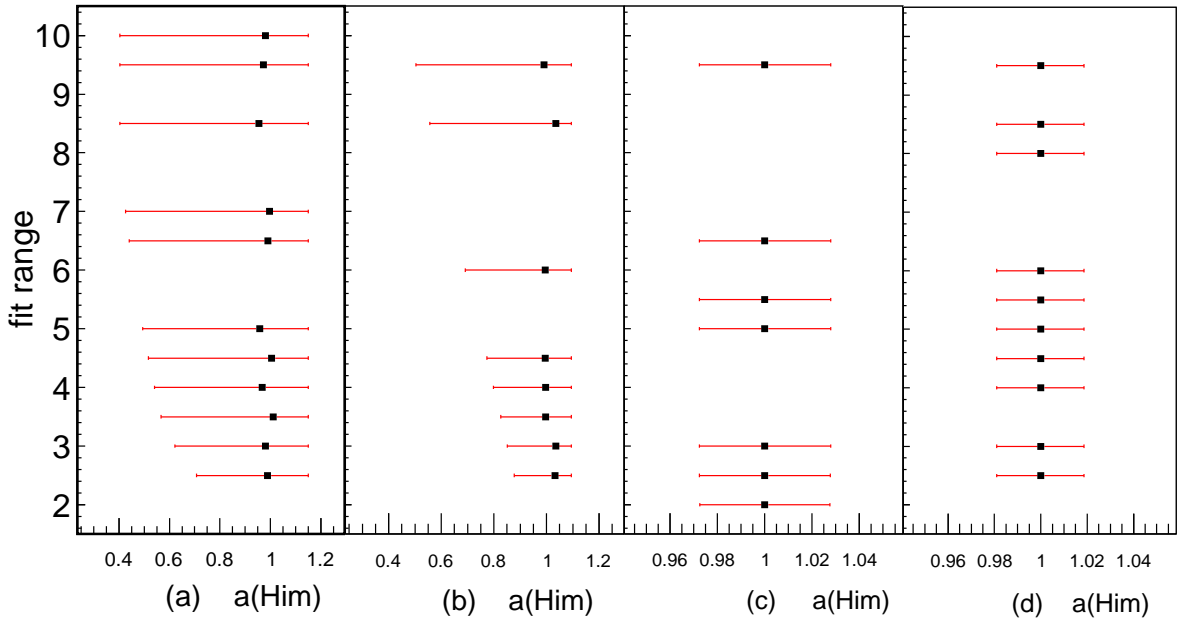}
\vspace{-2.cm}
\caption{Results of fits based on \underline{simulations} for the CFF multiplier $a(H_{Im})$ and its error bar ($x$-axis)
for different values of the boundaries for the domain of variation allowed for the CFFs ($y$-axis, in units of VGG CFFs) and for different sets
of observables fitted: (a) the unpolarized cross section $\sigma$ and the beam-polarized cross section $\Delta\sigma_{LU}$
(like the Hall A data), (b) $\sigma$, $\Delta\sigma_{LU}$ and $\Delta\sigma_{Uz}$, (c) $\sigma$, $\Delta\sigma_{LU}$, 
$\Delta\sigma_{Uz}$, $\Delta\sigma_{Ux}$ and $\Delta\sigma_{Uy}$, (d) $\sigma$ and all single and
double polarization observables. The pseudo-data that were fitted were generated with the VGG values, so that
the multipliers of all CFFs should be 1. 
In this example, the kinematics is approximately similar to the JLab Hall A data:
$\xi$=0.2, $Q^2$=3 GeV$^2$ and $-t$=0.4 GeV$^2$.}
\label{fig:simus}
\end{figure}

In order to give support to our results, we carried out simulations. We generated pseudo-data 
for all observables, associated with a 5\% error bar, from well-defined CFFs. We took as a simple example the VGG CFFs values so that 
the eight original $a(...)$ multipliers are equal to 1. Fig.~\ref{fig:simus} shows the results
of the fits for the multipliers, with their error bars, as a function of the maximum range
allowed for the variation of the CFFs: $\pm$ 2.5 times, $\pm$ 3 times, $\pm$ 3.5 times, ... the VGG CFFs. We made the study
for four different sets of observables fitted. When only the unpolarized cross section and the 
beam-polarized cross section are fitted (column (a) in Fig.~\ref{fig:simus}), i.e. the case of the
JLab Hall A data, the figure shows that: firstly, we are indeed able to recover the dominant $H_{Im}$ CFF that, along with
the 7 other CFFs, generated the pseudo-data, since $a(H_{Im})$ is essentially always very close to 1;
secondly, we find asymmetric error bars on $a(H_{Im})$, like we found for the fit of the real data;
and thirdly, the error bars on $a(H_{Im})$ increase when the allowed domain of variation of the CFFs increase.
These similar features between the fits of the data and of the simulations give a lot of confidence in the
method. We also learn from Fig.~\ref{fig:simus}, by comparing the four different sets of observables fitted, that the more observables we fit, the smaller are
the error bars on the resulting fitted CFF(s) and the more symmetric are these error bars. When we discuss and fit the HERMES data furtherdown, we will also observe this feature.  

To come back to real data, with the present method, we are therefore able to determine the $H_{Im}$ CFF for the four $t$-bins measured by the Hall A collaboration, as well as the $H_{Re}$ CFF for the two largest 
$t$-bins. In Table~\ref{tab:fitresult}, we also indicate the $\chi^2$ values of the fits for the 4 $t$-bins. One notices that the fit of the largest $-t$-bin ($-t$=0.33 GeV$^2$) is not ideal (the noramlized $\chi^2$ is around 3). This can be seen ``by eye" in the
right-most panel of Fig.~\ref{fig:halla_fit}. Therefore, the small error bars
obtained for $H_{Re}$ and $H_{Im}$ for this particular $t$-bin shall be taken with caution. It is not straightforward to interpret the error bar of a fitting parameter when the fit has a bad $\chi^2$.

\begin{figure}
\vspace{-6cm}
\begin{center}
\includegraphics[width =17.cm]{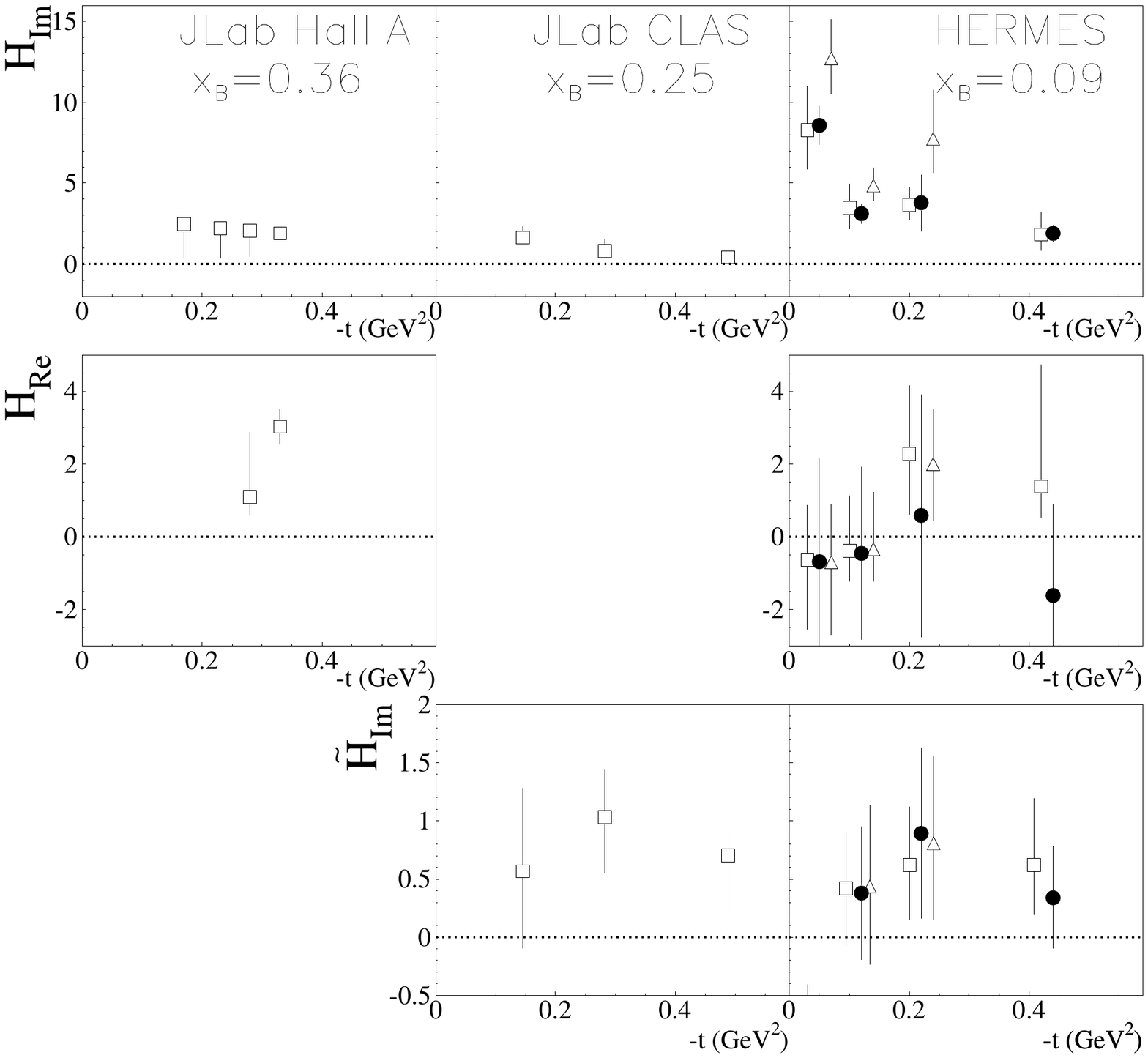}
\end{center}
%\vspace{-4cm}
\caption{The $H_{Im}$, $H_{Re}$ and $\tilde H_{Im}$ CFFs 
as a function of $-t$ for three different $x_B$ values. The empty squares 
show the results of the 8 CFFs fit presented in this article.
For the HERMES kinematics, the fits have been done with the newest data for the
beam spin asymmetry using the recoil detector~\cite{Airapetian:2012pg}. In empty triangles, we show the results of 
the 7 CFFs fit published in Ref.~\cite{Guidal:2010de} using the beam spin asymmetry data 
measured without this recoil detector. 
The solid circles show the result of the linear mapping fit of Ref.~\cite{Kumericki:2013lia}.
Points corresponding to the same $t$ value have been slightly shifted w.r.t. the
central $t$ value for visibility.}
\label{fig:allfit}
\end{figure}

We show in the upper left panel of Fig.~\ref{fig:allfit} with empty squares the values of the fitted $H_{Im}$ CFFs for the four $t$-bins of the JLab Hall A data. We recall that they result fom the multiplication of the fitted multipliers with the ``reference" VGG CFFs, all displayed in Table~\ref{tab:fitresult}. This result has been here updated 
from Ref.~\cite{Guidal:2008ie} due to the inclusion of the $E_{Im}$ CFF, i.e. an eighth free parameter, in the fit. The effect on the
final results is actually barely noticeable when one compares to the results of Ref.~\cite{Guidal:2008ie}.

In Fig.~\ref{fig:allfit}, we also display the updated results of our
fits (empty squares) for the CLAS and HERMES data. The CLAS data which consist of beam-polarized and longitudinally polarized target asymmetries allow to determine the $H_{Im}$ and $\tilde H_{Im}$ CFFs, at $<x_B>\approx 0.25$ for three $t$
-values. We recall that
  $\Delta\sigma_{LU}$, and therefore the associated asymmetry, is dominated by the
  $H_{Im}$ CFF (Eq.~(\ref{eq:relobscffa})) and that 
  $\Delta\sigma_{UL}$, and its associated asymmetry, is dominated by the
  $\tilde H_{Im}$ CFF (Eq.~(\ref{eq:relobscffb})). It is interesting to note that, for $H_{Im}$, although error bars are large and the kinematics are not exactly the same between the CLAS and the Hall A data, one extracts 
compatible values from fitting different observables (unpolarized and beam-polarized cross sections for Hall A and beam spin and longitudinally polarized target asymmetries
for CLAS) and that a better precision on the extraction 
of $H_{Im}$ is achieved by fitting two asymmetries than two cross sections.

The HERMES collaboration has measured an almost complete set of $ep\to e'p'\gamma'$ observables, for which only the unpolarized cross section is missing, at $<x_B>\approx 0.09$, and for four $t$-bins. This could allow in principle to determine all eight CFFs. However, in 
practice, the limited precision of the data allows to determine only three CFFs: $H_{Im}$, $H_{Re}$ and $\tilde H_{Im}$. The beam-polarized, beam-charge and longitudinally polarized target asymmetries are respectively dominated by these CFFs. In Fig.~\ref{fig:allfit}, we notice that the error bars on the fitted CFFs are, in the case of HERMES, essentially symmetric. As we saw in Fig.~\ref{fig:simus} for the case of many observables fitted simultaneously, simulations also showed this feature.
With so many constraints, the error bars on the fitted CFFs (of the order of 30\%) reflect more the statistics of the data than the correlations between the ``sub-dominant" and ``dominant" CFFs, in contrast with what we had for the Hall A and CLAS data cases. 

For this article, we have fitted the most recent data published by the HERMES collaboration for the beam-polarized asymmetry~\cite{Airapetian:2012pg}, where the recoil proton was detected with a recoil detector. For comparison we show in Fig.~\ref{fig:allfit}, with the empty triangles, the results of the fits
that we published earlier~\cite{Guidal:2010de} using the previous HERMES data, i.e. without the detection of the recoil proton,
and only 7 CFFs (no $E_{Im}$).
One can notice a systematic increase of the $H_{Im}$ CFFs when the new HERMES data are used. This directly reflects the
higher values (by about 30\%), for the four $t$-bins, of the new beam spin asymmetry moments measured by HERMES with the recoil detector. 
We also note that in the fit configuration of the present work, i.e. the inclusion
of the eighth CFF $E_{Im}$, some higher-twist corrections and the new HERMES data, the fit didn't converge
for the fourth $t$-bin and didn't allow to determine the three CFFs.
%-0.342 vs -0.225, -0.256 vs -0.193, -.409 vs -0.249 -0.336 vs -0.158 for

In Fig.~\ref{fig:allfit}, we also present the results of the alternative fit method of Ref.~\cite{Kumericki:2013lia} (solid circles).
This method consists in establishing a set of relations associating eight HERMES DVCS 
observables to the eight CFFs.
This is called ``mapping". 
Given some reasonable approximations (DVCS leading-twist and leading-order dominance,
neglect of some $\frac{t}{Q^2}$ terms in the analytical expressions, ...), this
set of relations can be linear. 
Eqs.~\ref{eq:relobscffa}-\ref{eq:relobscffd} give four examples of
such relations. The other ones can be found in Refs.~\cite{Belitsky:2001ns,Kumericki:2013lia}.
If a quasi-complete set of DVCS experimental observables can be measured
at a given ($x_B$, $-t$) point, one can then build a system of eight linear equations
with eight unknowns, \emph{i.e.} the eight CFFs, which can be
solved rather straightforwardly with standard matrix inversion and covariance error propagation techniques.

The HERMES data, which have the unique characteristic to consist of all beam-target single- 
and double-spin DVCS observables, along with beam-charge observables, readily lend themselves to this technique. The absence of cross section measurement at HERMES implies 
however that these spin observables are actually under the form of asymmetries, \emph{i.e.} a ratio of
polarized cross sections to unpolarized cross sections.
This has the consequence that the mapping is not fully linear and that some additionnal (reasonable)
approximations have to be made such as the dominance of the BH squared amplitude over
the DVCS squared amplitude.

With an appropriate selection (with some partial redefinition) of eight HERMES observables, 
the others serving as consistency checks, the authors of Ref.~\cite{Kumericki:2013lia} have 
thus been able to
solve the system of eight equations and extract the eight CFFs with their uncertaintities at the HERMES kinematics. As is the case with our method, with the actual precision of the HERMES data, only the $H_{Im}$, $\tilde H_{Im}$ and $H_{Re}$ come out with well-defined central values and errors, all others CFFs being compatible with zero within error bars. 

Fig.~\ref{fig:allfit} shows the results of this mapping technique for the three CFFs in the HERMES column with the solid circles. 
In this process, the HERMES beam spin asymmetry moments measured without the recoil detector were used.
They should therefore be compared with the results of our least-square 
minimization approach using
these same data, i.e. the empty triangles in Fig.~\ref{fig:allfit} (
we recall that the empty squares used the new HERMES data \underline{with} the recoil detector).
The general agreement between the mapping technique and our method is,
for the central values of the three CFFs $H_{Im}$ $H_{Re}$ and $\tilde H_{Im}$, remarkable. 

We conclude this section by a rapid physics discussion of the first hints on nucleon structure that can be drawn from these CFF fits. In Fig.~\ref{fig:allfit}, although error bars are large, some general features and trends can be distinguished:
\begin{itemize}
\item Concerning $H_{Im}$, it appears that, at (approximately) fixed $-t$, 
it increases as $x_B$ decreases, \emph{i.e.} going from JLab to HERMES kinematics. Since the GPD $H$ reflects the helicity-independent quark densities in an un polarized nucleon, this rise reflects, like for standard unpolarized parton distribution (to which $H_{Im}$ reduces at $\xi$=0 and $t=0$), the density of quarks (and anti-quarks) rising as smaller (longitudinal) momentum fractions are probed. This rise is essentially due to the sea quarks (and anti-quarks). 
\item  Another feature concerning $H_{Im}$ is that its $t-$slope appears to increase 
with $x_B$ decreasing. We recall that the $t$-slope of the
GPD is related, via a Fourier transform and some ``deskewing" effect 
(GPD$(\xi,\xi,t)\to$$GPD(\xi,0,t)$) which was estimated in Ref.~\cite{Guidal:2013rya}, 
to the transverse spatial densities of quarks in the nucleon.
This evolution of the $t$-slope with $x_B$ suggests then that low-$x$ quarks (the ``sea") 
would extend to the periphery of the nucleon while the high-$x$ (the 
``valence") would tend to remain in the center of the nucleon. 
\item Concerning $\tilde H_{Im}$, which reflects the helicity-dependent quark densities in the nucleon, we notice that it is in general
smaller than $H_{Im}$, which can be expected for a polarized
quantity compared to an unpolarized one. 
We also observe that the $t$-dependence of $\tilde H_{Im}$ is rather flat. This weaker $t$-dependence 
of $\tilde H_{Im}$ compared to $H_{Im}$ suggests
that the axial charge (to which the $\tilde{H}$ GPD relates at $\xi$=0 and $t=0$) is more concentrated in the nucleon than the electromagnetic charge. We remark that
the slope of the axial FF is also well known to be flatter compared to the one of the Dirac electromagnetic FF, which points to the same feature.
It is quite comforting that by studying two relatively different experimental 
processes (DVCS and for instance $\pi$-electroproduction from which the axial FF is in general
extracted), one finds similar conclusions. One can also note that there is very little $x_B$ dependence 
for $\tilde H_{Im}$ which suggests that this concentration of the axial charge would remain for all quarks' momentum. 

\end{itemize}

\section{Conclusion}

In this review, we have shown how to extract some information from
fitting an under-constrained system, taking advantage of the fact that:
1/a few equations (observables) are dominated by a few parameters (CFFs)
and 2/the range of variation of the parameters can be limited (in
a conservative way). The latter is esssentially the only model-dependent
input in the problem. Our approach was supported by simulations 
which displayed similar feature as for real data and our
results found in agreement with an 
alternative independent approach~\cite{Kumericki:2013lia} 
(for the HERMES case for which a sufficient number of observables is available).

With the currently (scarce) existing data and the results of our method, some first 
physics hints on nucleon
structure come out, in particular the way quarks are distributed (in space and in momentum)
inside the nucleon can already be drawn, i.e. the way the nucleon
transverse size increases as lower momentum quarks are probed.

A wealth of new DVCS-related data is expected in the near future
from JLab (in particular with the forthcoming 12 GeV beam energy upgrade)
and COMPASS, bringing new observables and more precise data. It can be
shown with simulations~\cite{Guidal:2013rya} that with these projected
data, all CFFs will be extracted with good accuracy. Getting from CFFs to GPDs
requires still more efforts. We are progressing step by step and we have presented in this article one of the first tools developped paving the way towards this GPD quest.

\section*{Acknowledgements}

This work is partly supported by the Joint Research Activity ``GPDex"
of the European program Hadron Physics 3 under the Seventh
Framework Programme of the European Community, by the French GDR 3034
``PH-QCD" and the ANR-12-MONU-0008-01 ``PARTONS". Also, the
simulations presented in this article have been carried out
with the grid infrastructure ``GRIF" (http://www.grif.fr/) and
we are very thankful to C. Diarra and C. Suire for the support.

\end{document}